\documentclass[preprintnumbers,nofootinbib,superscriptaddress,10pt]{revtex4}
\usepackage[dvipdfmx]{graphicx} 
\usepackage{amsmath,amssymb,amsfonts,bm,comment} 

\def\a{\alpha}  \def\g{\gamma}  \def\d{\delta}          \def\m{\mu}         \def\t{\tau}       

  \def\nn{\nonumber}

\newcommand{\DDiag}[4]{ 
\begin{pmatrix}
 #1 & 0 & 0 & 0 \\ 0 & #2 & 0 & 0 \\ 0 & 0 & #3  &0 \\ 0 & 0 &0 & #4 \\
\end{pmatrix}
}

\usepackage{color}

\begin{document}

\title{\large Explicit rephasing transformation of four-generation mixing matrix and formulae for CP phases}

\preprint{STUPP-25-290}

\author{Masaki J. S. Yang}
\email{mjsyang@mail.saitama-u.ac.jp}
\affiliation{Department of Physics, Saitama University, 
Shimo-okubo, Sakura-ku, Saitama, 338-8570, Japan}
\affiliation{Department of Physics, Graduate School of Engineering Science,
Yokohama National University, Yokohama, 240-8501, Japan}



\begin{abstract} 

In this letter, we present an explicit rephasing transformation that maps a general $4\times4$ flavor mixing matrix 
to the standard parametrization of four-generation models.  
By combining rephasing-covariant minors and the determinant systematically, we derive expressions for all three Dirac-type CP phases, three Majorana phases, and four unphysical phases by arguments of matrix elements.
The resulting formulae smoothly reduce to the three-generation results in the limit where the fourth generation decouples.

\end{abstract} 

\maketitle

\section{Introduction}

The fourth generation of fermions is one of the possible extensions of the Standard Model \cite{Barger:1984jc, Marciano:1985pd}.
In the context of Higgs physics, the contributions of a fourth generation and its viability continue to be examined through experimental and phenomenological studies \cite{Kribs:2007nz, Eberhardt:2012gv}.
In neutrino physics, the possibility of a fourth neutrino has been discussed \cite{Gibbons:1998pg,Barger:1998bn,Barger:1999hi,Xing:2001bg,Guo:2001yt}, and this line of investigation has naturally evolved into the current research on sterile neutrinos \cite{Karagiorgi:2006jf,Klop:2014ima, Gandhi:2015xza, Cianci:2017okw, Choubey:2017ppj, Xing:2020ivm}.

With the fourth generation, the flavor mixing matrix is extended to a 
$4 \times 4$ unitary matrix, and the number of physical CP-violating phases increases to three \cite{Botella:1985gb}.
The phase structure of the mixing matrix becomes more involved than in the three-generation case, requiring a more systematic treatment for its analysis.
These CP phases have been studied through various rephasing invariants, most notably the Jarlskog invariant and its generalizations \cite{Jarlskog:1985ht, Wu:1985ea, Bernabeu:1986fc, Gronau:1986xb, Branco:1987mj, Bjorken:1987tr, Nieves:1987pp, Botella:1994cs, Kuo:2005pf, Jenkins:2007ip, Chiu:2015ega}.
In such approaches, the magnitude of CP violation is expressed by the imaginary parts of these invariants, through $\sin \d$-type factors. However, many of these studies did not directly elucidate the phase structure of the mixing matrix itself.

Recently,  the Dirac CP phase in the PDG parametrization is expressed directly in terms of arguments of the matrix elements and the determinant \cite{Yang:2025hex,Yang:2025cya,Yang:2025law,Yang:2025ftl,Yang:2025dhm,Yang:2025vrs}.
Although the determinant represents an unphysical overall phase, 
the information contained in the determinant is indispensable for a complete description of CP phases. 
Furthermore, a previous work \cite{Yang:2025dkm} shows an explicit construction of the rephasing transformation that maps a mixing matrix in an arbitrary basis to the standard PDG parametrization. 

The purpose of this study is to extend this framework to mixing matrices with a fourth generation and to construct explicit rephasing transformations. 
This allows us to derive transparent formulae that express all three Dirac-type CP phases, three Majorana phases, and four unphysical phases. 

\section{Explicit Rephasing Transformation to the Standard Form of a Four-Generation Mixing Matrix}

We define the standard form of the mixing matrix $U_{\rm std} = U^{0} P$ for four generations as \cite{Botella:1985gb}
\begin{align}
U^{0} = R_{34} \tilde R_{24} \tilde R_{14} R_{23} \tilde R_{13} R_{12} = 
\begin{pmatrix}
c_{12} c_{13} c_{14} & c_{13} c_{14} s_{12} & c_{14} s_{13} e^{-i \delta _{e3} } & s_{14} e^{-i \delta _{e4} } \\
U_{\m 1}^{0} & U_{\m 2}^{0} & U_{\m 3}^{0} & c_{14} s_{24} e^{-i \delta _{ \mu 4} } \\
 U_{\t 1}^{0} & U_{\t 2}^{0} & U_{\t 3}^{0} & c_{14} c_{24} s_{34} \\
 U_{s 1}^{0} & U_{s2}^{0} & U_{s 3}^{0} & c_{14} c_{24} c_{34} \\
\end{pmatrix} ,  \nn
\end{align}
where $R_{ij}$ denotes a real $2\times2$ rotation in the $ij$ block, and
 $\tilde R_{ij}$ denotes a rotation that contains a CP-violating phase $\d_{\a j}$.
The matrix elements required for the subsequent calculations are
\begin{align}
U_{\m 1}^{0} & =  -c_{23} c_{24} s_{12}+c_{12} \left( - c_{24} s_{13} s_{23} e^{i \delta _{e3} } -c_{13} s_{14} s_{24} e^{i \delta _{e4} - i \delta _{\mu 4} }\right)  \, , \\
U_{\m 2}^{0} & = c_{12} c_{23} c_{24}+s_{12} \left(- c_{24} s_{13} s_{23} e^{i \delta _{e3} } -c_{13} s_{14} s_{24} e^{i \delta _{e4} - i \delta _{ \mu 4 } }\right) \, , \\
U_{\m 3}^{0} & = c_{13} c_{24} s_{23}-s_{13} s_{14} s_{24} e^{-i \delta _{\text{$\mu $4}}-i \delta _{ e3 } + 
i \delta _{e4 } } \, , \\ 
U_{\t 3}^{0} &=c_{13} \left(c_{23} c_{34}-s_{23} s_{24} s_{34} e^{i \delta _{ \mu 4} } \right)-c_{24} s_{13} s_{14} s_{34} e^{i \delta _{e4} - i \delta _{ e3} }  \, , \\
U_{s 3}^{0} & = c_{13} \left(-c_{23} s_{34} - c_{34} s_{23} s_{24} e^{i \delta _{ \mu 4} } \right)-c_{24} c_{34} s_{13} s_{14} e^{i \delta _{e4} - i \delta _{e3} }  \, . 
\end{align}
The remaining matrix elements in the lower-left block will be expressed later in terms of the arguments of these matrix elements.
The Majorana phases are defined to follow the standard PDG convention \cite{ParticleDataGroup:2018ovx} 
\begin{align}
P = {\rm diag	} ( 1 \, , \, e^{i \a_{2} /2}\, , \, e^{i \a_{3} /2}\, , \, e^{i \a_{4} /2} ) \, . 
\end{align}

We perform an explicit rephasing transformation by deriving the CP phases $\d_{\a i}$ and the Majorana phases $\a_{i}$ through rephasing transformations. 
Let us consider a mixing matrix $U$ in an arbitrary basis, 
which is transformed from $U^{0}$ by a rephasing transformation 
\begin{align}
U = 
\DDiag{e^{i \g_{L1}}}{e^{i \g_{L2}}}{e^{i \g_{L3}}}{e^{i \g_{L4}}}
\begin{pmatrix}
|U_{e1}| & |U_{e2}| & |U_{e3}| e^{-i\d_{e3}} & |U_{e4}|  e^{-i\d_{e4}}  \\
U_{\m1}^{0} & U_{\m 2}^{0} & U_{\m 3}^{0} & |U_{\m4}|  e^{-i\d_{\m 4}}   \\
U_{\t 1}^{0} & U_{\t 2}^{0} & U_{\t 3}^{0} & |U_{\t 4}|   \\
U_{s 1}^{0} & U_{s 2}^{0} & U_{s 3}^{0} & |U_{s 4}|   \\
\end{pmatrix}
\DDiag{e^{ - i \g_{R1}}}{e^{ - i \g_{R2}}}{e^{ - i \g_{R3}}}{e^{ - i \g_{R4}}} .
\end{align}
These phases of the redefinition $\g_{L i}$ and $\g_{R i}$ will be  expressed in terms of arguments of the matrix elements. 
It is immediately seen that
\begin{align}
& \arg U_{e1} = \g_{L1} - \g_{R1} \, , ~
\arg U_{e2} = \g_{L1} - \g_{R2} \, , ~
\arg U_{e3} = \g_{L1} - \g_{R3} - \d_{e 3}  \, , ~
\arg U_{e 4} = \g_{L1} - \g_{R4} - \d_{e4} \, , \nn  \\
&  \arg U_{\m 4} = \g_{L2} - \g_{R 4} - \d_{\m 4} \, , ~ 
\arg U_{\t 4} = \g_{L3} - \g_{R 4} \, , ~ 
\arg U_{s 4} = \g_{L4} - \g_{R 4} \, . 
\end{align}
Reflecting the freedom of an overall phase, one of the eight phases $\g_{L i}$ and $\g_{R i}$ cannot be determined. 
To preserve the form of Majorana phases, we leave $\g_{R1}$ as an unknown variable and solve the remaining phases accordingly.

The phases are solved by using the arguments of second-order minor determinants.
We first consider the following minor in the standard form, 
\begin{align}
U_{e1}^{0} U_{\m 2}^{0} - U_{e2}^{0} U_{\m1}^{0} = c_{13} c_{14} c_{23} c_{24} \, . 
\end{align}
Since this minor has no CP phase in the standard form, it transforms covariantly under rephasing transformations, 
\begin{align}
\arg [ U_{e1} U_{\m 2} - U_{e2} U_{\m1}  ] =
\arg [ {U_{e1} U_{\m 2} - U_{e2} U_{\m1} \over U_{e1}^{0} U_{\m 2}^{0} - U_{e2}^{0} U_{\m1}^{0}} ] = 
 \g_{L1} + \g_{L2} - \g_{R1} - \g_{R2} \, . 
\end{align}
From this, the phases $\g_{L1}$, $\g_{L2}$, and $\g_{R2}$ in the upper-left $2\times2$ block are determined as 
\begin{align}
\g_{L1} & = \arg U_{e1} + \g_{R1} \, , ~~ 
\g_{R2} = \arg \left[ {U_{e1} \over U_{e2} } \right] + \g_{R1} \, , ~~ 
\g_{L2} = \arg \left[ { U_{e1} U_{\m 2} - U_{e2} U_{\m1} \over  U_{e2} } \right]  + \g_{R1} \, . 
\end{align}

Similarly, since the following rephasing covariant is real in the standard form $\tilde U_{\m3}^{0} = c_{14}^{2} c_{12} c_{24} c_{23}$, 
\begin{align}
\tilde U_{\m 3} & \equiv (1 - |U_{e4}|^{2} ) U_{\m 3} + U_{e 3} U_{e4}^{*} U_{\m4} 
 =  U_{\m 3} + U_{e4}^{*} ( U_{e 3} U_{\m4} - U_{e4}  U_{\m 3} ) \, , 
\end{align}
this $\tilde U_{\m 3}$ carry another phase difference 
\begin{align}
\arg \tilde U_{\m 3} = 
\arg [ \tilde U_{\m 3} / \tilde U_{\m 3}^{0} ]= \g_{L2} - \g_{R3} \, . 
\end{align}
It allows $\g_{R3}$ to be solved as 
\begin{align}
\g_{R3} = \g_{L2} - \arg \tilde U_{\m 3}  = \arg \left[ {U_{e1} U_{\m2} - U_{e2} U_{\m 1} \over U_{e2} \tilde U_{\m 3} } \right ] + \g_{R1} \, ,
\end{align}
and the first CP phase $\d_{e3}$ is obtained as an argument of a rephasing ivariant 
\begin{align}
\d_{e3} = \arg [U_{e1} / U_{e3}] - \g_{R3} + \g_{R1} 
= \arg \left[ {  U_{\m 3} + U_{e4}^{*} ( U_{e 3} U_{\m4} - U_{e4}  U_{\m 3} ) \over U_{e1}^{*} U_{e2}^{*} U_{e3} (U_{e1} U_{\m2} - U_{e2} U_{\m 1} ) } \right ] . 
\end{align}
When the fourth generation decouples, the minor $(U_{e1} U_{\m2} - U_{e2} U_{\m 1}) $ reduces to the element of the inverse matrix
$U_{\t 3}^{*} \det U$, and the CP phase reduces to the three-generation result 
$\d = \arg \left[ { U_{e1} U_{e2} U_{\m 3} U_{\t 3} / U_{e3} \det U } \right ]$.

Finally, by focusing on the lower-right $2\times2$ block, 
the remaining phases $\g_{L3}$, $\g_{L4}$, and $\g_{R4}$ are 
\begin{align}
 \g_{R4} = \g_{L4} - \arg U_{s 4}  \, , ~ 
 \g_{L3} = \arg U_{\t 4} + \g_{R 4} = \arg [U_{\t 4} /U_{s 4} ] + \g_{L 4} \, .
\end{align}
The following minor carries no CP phase in the standard form again,
\begin{align}
 & \arg [U_{\t 3} U_{s 4} - U_{\t 4} U_{s 3} ]  = \g_{L3} + \g_{L4} - \g_{R3} - \g_{R4} \, ,  \\
 \g_{L4} & = \arg \left[{U_{e1} U_{\m2} - U_{e2} U_{\m 1} \over U_{e2} \tilde U_{\m 3} }  { U_{\t 3} U_{s 4} - U_{\t 4} U_{s 3} \over U_{\t 4} } \right ] + \g_{R1}  \, . 
\end{align}
Since the following rephasing invariant does not have a complex phase,
\begin{align}
\arg \left[ { \det U \over (U_{e1} U_{\m 2} - U_{e2} U_{\m 1} )(U_{\t 3} U_{s 4} - U_{\t 4} U_{s 3}) } \right ] = 0 \, , 
\end{align}
the remaining phases are immediately determined as
\begin{align}
\g_{L 4} = \arg  \left[ {  \det U  \over U_{e2} \tilde U_{\m 3} U_{\t 4} } \right ] + \g_{R 1} \ ,  ~
\g_{R 4} = \arg  \left[ {  \det U  \over U_{e2} \tilde U_{\m 3} U_{\t 4} U_{s 4} } \right ] + \g_{R 1}  \, ,  ~ 
 \g_{L3} = \arg  \left[ {  \det U  \over U_{e2} \tilde U_{\m 3} U_{s 4} } \right ] + \g_{R 1}  \, . 
\end{align}
The second CP phase is then obtained from $\arg U_{\m 4} = \g_{L2} - \g_{R4} - \d_{\m4}$, 
\begin{align}
\d_{\m 4} & = \arg \left[ { \tilde U_{\m 3}  U_{\t4}  U_{s 4}  \over U_{\m 4}  (U_{\t 3} U_{s 4} - U_{\t 4} U_{s 3}) } \right ]
= \arg \left[ { U_{\m 3} + U_{e4}^{*} ( U_{e 3} U_{\m4} - U_{e4}  U_{\m 3} ) \over U_{\m 4} U_{\t4}^{*}  U_{s 4}^{*} (U_{\t 3} U_{s 4} - U_{\t 4} U_{s 3}) } \right ]  . 
\end{align}
Since this has a symmetric form to $\d_{e3}$, it similarly reduces to the three-generation result when the first generation decouples.
Finally, the last CP phase is obtained from $\arg U_{e 4} = \g_{L1} - \g_{R4} - \d_{e4}$ as 
\begin{align}
\d_{e4} 
= \arg \left[ { U_{e1} U_{e2} \tilde U_{\m 3} U_{\t 4} U_{s4} \over U_{e 4} \det U } \right ] 
= \arg \left[ { U_{\m 3} + U_{e4}^{*} ( U_{e 3} U_{\m4} - U_{e4}  U_{\m 3} )  \over U_{e1}^{*} U_{e2}^{*} U_{e 4} U_{\t 4}^{*} U_{s4}^{*} \det U } \right ]  . 
\end{align}
This completes the determination of all ten phases.

As a result, the rephasing transformation of a four-generation mixing matrix $U$ 
to the standard form is  performed explicitly 
\begin{align}
U & = 
(e^{i \arg U_{e1} } \, , \, 
e^{ i  \arg \left[ { U_{e1} U_{\m2} - U_{e2} U_{\m1} \over U_{e2} } \right] } \, , \,
e^{ i \arg \left[ {  \det U \over U_{e2} \tilde U_{\m3} U_{s4} } \right] } \, , \,
e^{ i \arg \left[ {  \det U \over U_{e2} \tilde U_{\m3} U_{\t 4}  } \right] }  ) \nn \\
& \times U^{0}
(1 \, , \, 
e^{ i \arg [{U_{e2}\over U_{e1} }] } \, , \, 
e^{  i \arg \left[ {U_{e2} \tilde U_{\m 3} \over U_{e1} U_{\m2} - U_{e2} U_{\m 1}  } \right ] } \, , \, 
e^{ i \arg  \left[ { U_{e2} \tilde U_{\m 3} U_{\t 4} U_{s 4} \over  \det U  } \right ] }  ) \, . 
\end{align}
The unknown phase $\g_{R1}$ cancels between the left and right phase matrices, and 
other right-hand phases $\g_{R i}$ directly yield the Majorana phases 
\begin{align}	
{\a_{2}\over 2} = \arg \left[ {U_{e2}\over U_{e1} } \right] , ~~
{\a_{3} \over 2} = \arg \left[ {U_{e2} \tilde U_{\m 3} \over U_{e1} U_{\m2} - U_{e2} U_{\m 1}  } \right ] , ~~ 
{\a_{4} \over 2} = \arg  \left[ { U_{e2} \tilde U_{\m 3} U_{\t 4} U_{s 4} \over  \det U  } \right ] .  
\end{align}
The remaining four $\g_{L i}$ are unphysical phases.

As a consistency check, 
the sums of the left- and right-hand phases cancel pairwise, resulting in the argument of the determinant 
\begin{align}
\sum_{i} \g_{Li} - \g_{Ri} = \arg \det U \, . 
\end{align}
Since the following product of the elements with the CP phases transforms in the same way as $\det U$,
\begin{align}
 \arg \left[ { U_{e1} U_{e2}U_{e3}  U_{\m 4}  U_{\t 4} U_{s4} / U_{e4}^{2} } \right] + \d_{e3} + \d_{\m 4}  - 2 \d_{e4} 
 = \sum_{i} \g_{Li} - \g_{Ri}  =  \arg \det U  \, , 
\end{align}
we obtain a sum rule relating the CP phases and the matrix elements 
\begin{align}
 2 \d_{e 4} - \d_{e3} - \d_{\m 4}  = \arg \left[ { U_{e1} U_{e2} U_{e3} U_{\m 4}  U_{\t 4} U_{s 4} \over U_{e4}^{2} \det U } \right]  . 
\end{align}
The sum rule is verified directly from the three solutions. 
Moreover, by applying the same reasoning to the other two minors,
\begin{align}
\d_{\m 4} - \d_{e4} = \arg \left[ { U_{e4} ( U_{e1} U_{\m 2} - U_{e2} U_{\m1}) \over U_{e1} U_{e2}  U_{\m 4} } \right] , ~~ 
\d_{e3} - \d_{e4} = \arg \left[ {U_{e4}(U_{\t 3} U_{s 4} - U_{\t 4} U_{s 3})  \over  U_{e3}  U_{\t 4} U_{s 4} } \right] . 
\end{align}
These expressions are more concise because $\tilde U_{\m 3}$ does not appear.
Various sum rules among the CP phases and other invariants are expected to hold, in close analogy with the three-generation case.

Since a $4 \times 4$ unitary matrix has only ten phases, 
the six nontrivial arguments are not independent and are expressed by the remaining phases.
Through this explicit rephasing transformation, one finds the following relations
 between the matrix elements in an arbitrary basis and those in the standard form 
\begin{align}
U_{\m1} & = e^{ i  \arg \left[ { U_{e1} U_{\m2} - U_{e2} U_{\m1} \over U_{e2} } \right] }  U_{\m 1}^{0} \, , ~~
 U_{\m 2} = e^{ i  \arg \left[ { U_{e1} U_{\m2} - U_{e2} U_{\m1} \over U_{e1} } \right] } U_{\m 2}^{0}  \, , ~~
 U_{\m 3} = e^{i \arg [ \tilde U_{\m3} ] } U_{\m 3}^{0} \, , \\
 U_{\t 1} & = e^{ i \arg \left[ {  \det U \over U_{e2} \tilde U_{\m3} U_{s4} } \right] } U_{\t 1}^{0}\, , ~~
 U_{\t 2}  =  e^{ i \arg \left[ {  \det U \over U_{e1} \tilde U_{\m3} U_{s4} } \right] } U_{\t 2}^{0}\, ,  ~~
 U_{\t 3}  = e^{ i  \arg \left[ { U_{\t 3} U_{s 4} - U_{\t 4} U_{s 3} \over U_{s 4} } \right] } U_{\t 3}^{0} \, , \\
 U_{s 1} & =  e^{ i \arg \left[ {  \det U \over U_{e2} \tilde U_{\m3} U_{\t 4}  } \right] }  U_{s 1}^{0} \, , ~~
 U_{s 2}  =  e^{ i \arg \left[ {  \det U \over U_{e1} \tilde U_{\m3} U_{\t 4}  } \right] }  U_{s 2}^{0} \, , ~~ 
 U_{s 3}  = e^{ i  \arg \left[ { U_{\t 3} U_{s 4} - U_{\t 4} U_{s 3} \over U_{\t 4} } \right] } U_{\t 3}^{0}  \, . 
\end{align}
In particular, the four arguments in the lower-left block are expressed by other rephasing invariants
\begin{align}
\arg U_{\t 1}^{0} & = \arg \left[ { U_{e2} \tilde U_{\m3} U_{\t1} U_{s4} \over  \det U } \right] \, , ~~ 
\arg U_{\t 2}^{0}  = \arg \left[ { U_{e1} \tilde  U_{\m3} U_{\t2} U_{s4} \over  \det U } \right] \, , \\
\arg U_{s 1}^{0} & = \arg \left[ { U_{e2} \tilde  U_{\m3} U_{\t 4} U_{s1} \over \det U } \right] \, , ~~ 
\arg U_{s 2}^{0} = \arg \left[ { U_{e1} \tilde  U_{\m3} U_{\t 4} U_{s2} \over \det U } \right] \, .
\end{align}
In the standard parametrization, $U_{e1}^{0}$, $U_{e2}^{0}$, $\tilde U_{\m 3}^{0}$, $U_{\t 4}^{0}$, and $U_{s4}^{0}$ carry only trivial phases, and  the validity of these relations is evident.
Since these four matrix elements are highly nontrivial trigonometric functions,
the resulting compact expressions are expected to greatly simplify studies of the fourth generation and sterile neutrinos.

\section{Summary}

In this letter, we have constructed an explicit rephasing transformation to the standard parametrization for a general $4\times4$ flavor mixing matrix of the four-generation models.
By combining rephasing-covariant minors and the determinant systematically, 
all three Dirac-type CP phases, three Majorana phases, and four unphysical phases are expressed in terms of the arguments of the mixing matrix elements.
The resulting formulae for the CP phases naturally reduce to the three-generation results in the limit where the fourth generation decouples.
This explicit rephasing is applicable not only to the mixing matrix itself, but also directly to the diagonalization matrices of fermion mass matrices $U_{f}$. 
These results are expected to be useful for analyzing the more fundamental origins of phases, and 
will provide a practical and powerful tool in studies of CP violation in fourth-generation models and sterile-neutrino scenarios. 

\section*{Acknowledgment}

The study is partly supported by the MEXT Leading Initiative for Excellent Young Researchers Grant Number JP2023L0013.


\end{document}